\documentclass[twocolumn,showpacs,preprintnumbers,amsmath,amssymb,superscriptaddress]{revtex4}

\usepackage{graphicx,epsfig}
\usepackage{dcolumn}
\usepackage{bm}

\begin{document}

\title{ Further search for the decay $K^+\to \pi^+ \nu \bar\nu$ in the 
momentum region $P < 195 {\rm ~MeV/c}$}

\author{S.~Adler} \affiliation{Brookhaven National Laboratory, Upton,
New York 11973} 

\author{M.~Aoki}
\altaffiliation{Present address: Department of Physics, Osaka
University, Toyonaka, Osaka 560-0043, Japan.} 
\affiliation{TRIUMF, 4004 Wesbrook Mall, Vancouver, British Columbia,
Canada, V6T 2A3} 

\author{M.~Ardebili} \affiliation{Joseph Henry
Laboratories, Princeton University, Princeton, New Jersey 08544}

\author{M.S.~Atiya} \affiliation{Brookhaven National Laboratory, Upton,
New York 11973} 

\author{A.O.~Bazarko} \affiliation{Joseph Henry
Laboratories, Princeton University, Princeton, New Jersey 08544}

\author{P.C.~Bergbusch} \affiliation{Department of Physics and
Astronomy, University of British Columbia, Vancouver, British
Columbia, Canada, V6T 1Z1}

\author{B.~Bhuyan} \altaffiliation{Also at Department of Physics and Astrophysics, 
University of Delhi, Delhi, India. Present address: 
Department of Physics and Astronomy, 
University of Victoria, Victoria, BC, Canada V8W 3P6}  
\affiliation{Brookhaven National Laboratory, Upton,
New York 11973} 

\author{E.W.~Blackmore}
\affiliation{TRIUMF, 4004 Wesbrook Mall, Vancouver, British Columbia,
Canada, V6T 2A3} 

\author{D.A.~Bryman} \affiliation{Department of
Physics and Astronomy, University of British Columbia, Vancouver,
British Columbia, Canada, V6T 1Z1}

\author{I-H.~Chiang} \affiliation{Brookhaven National
Laboratory, Upton, New York 11973}

\author{M.R.~Convery} \altaffiliation{Present address: 
Stanford Linear Accelerator Center, Stanford, CA 94309}  
\affiliation{Joseph Henry
Laboratories, Princeton University, Princeton, New Jersey 08544}

\author{M.V.~Diwan}
\affiliation{Brookhaven National Laboratory, Upton, New York 11973}

\author{J.S.~Frank} \affiliation{Brookhaven National Laboratory,
Upton, New York 11973} 

\author{J.S.~Haggerty} \affiliation{Brookhaven
National Laboratory, Upton, New York 11973}

\author{T.~Inagaki} \affiliation{High Energy
Accelerator Research Organization (KEK), Oho, Tsukuba, Ibaraki
305-0801, Japan} 

\author{M.M.~Ito}\altaffiliation{Present address:
Thomas Jefferson
National Accelerator Facility, Newport News, Virginia 23606.}
\affiliation{Joseph Henry Laboratories,
Princeton University, Princeton, New Jersey 08544}

\author{V.~Jain}
\affiliation{Brookhaven National Laboratory, Upton, New York 11973}

\author{D.E.~Jaffe}
\affiliation{Brookhaven National Laboratory, Upton, New York 11973}

\author{S.~Kabe} \affiliation{High Energy Accelerator Research
Organization (KEK), Oho, Tsukuba, Ibaraki 305-0801, Japan}

\author{M.~Kazumori} \affiliation{High Energy
Accelerator Research Organization (KEK), Oho, Tsukuba, Ibaraki
305-0801, Japan} 

\author{S.H.~Kettell} \affiliation{Brookhaven National Laboratory,
Upton, New York 11973}
 
\author{P.~Kitching} \affiliation{Centre for
Subatomic Research, University of Alberta, Edmonton, Canada, T6G 2N5}

\author{M.~Kobayashi} \affiliation{High Energy Accelerator Research
Organization (KEK), Oho, Tsukuba, Ibaraki 305-0801, Japan}

\author{T.K.~Komatsubara} \affiliation{High Energy Accelerator
Research Organization (KEK), Oho, Tsukuba, Ibaraki 305-0801, Japan}

\author{A.~Konaka} \affiliation{TRIUMF, 4004 Wesbrook Mall, Vancouver,
British Columbia, Canada, V6T 2A3}

\author{Y.~Kuno}
\altaffiliation{Present address: Department of Physics, Osaka
University, Toyonaka, Osaka 560-0043, Japan.} 

 \affiliation{High
Energy Accelerator Research Organization (KEK), Oho, Tsukuba, Ibaraki
305-0801, Japan} 

\author{M.~Kuriki} 
\affiliation{High Energy
Accelerator Research Organization (KEK), Oho, Tsukuba, Ibaraki
305-0801, Japan} 

\author{T.F.~Kycia} \altaffiliation{Deceased}\affiliation{Brookhaven 
National Laboratory, Upton, New York 11973}

\author{K.K.~Li} \affiliation{Brookhaven National Laboratory,
Upton, New York 11973}

\author{L.S.~Littenberg} \affiliation{Brookhaven National Laboratory,
Upton, New York 11973} 

\author{J.A.~Macdonald} \altaffiliation{Deceased}\affiliation{TRIUMF,
4004 Wesbrook Mall, Vancouver, British Columbia, Canada, V6T 2A3}

\author{R.A.~McPherson} \altaffiliation{Present address: 
Department of Physics and Astronomy, 
University of Victoria, Victoria, BC Canada}  
\affiliation{Joseph
Henry Laboratories, Princeton University, Princeton, New Jersey 08544}

\author{ P.D.~Meyers} \affiliation{Joseph Henry Laboratories,
Princeton University, Princeton, New Jersey 08544}

\author{J.~Mildenberger} \affiliation{TRIUMF, 4004 Wesbrook Mall,
Vancouver, British Columbia, Canada, V6T 2A3}

\author{N.~Muramatsu}
\altaffiliation{Present address: Department of Physics, Kyoto University,
Sakyo-ku, Kyoto 606-8502, Japan}
\affiliation{Research Center for Nuclear Physics, Osaka University,
10-1 Mihogaoka, Ibaraki, Osaka 567-0047, Japan}

\author{T.~Nakano}
\affiliation{Research Center for Nuclear Physics, Osaka University,
10-1 Mihogaoka, Ibaraki, Osaka 567-0047, Japan} 

\author{C.~Ng}
\altaffiliation{Also at Physics Department, State University of New
York at Stony Brook, Stony Brook, NY 11794-3800.}
\affiliation{Brookhaven National Laboratory, Upton, New York 11973}

\author{S.~Ng} \affiliation{Centre for
Subatomic Research, University of Alberta, Edmonton, Canada, T6G 2N5}

\author{T.~Numao} \affiliation{TRIUMF, 4004 Wesbrook Mall, Vancouver,
British Columbia, Canada, V6T 2A3}

\author{A.~Otomo} 
\affiliation{High Energy
Accelerator Research Organization (KEK), Oho, Tsukuba, Ibaraki
305-0801, Japan}

\author{J.-M.~Poutissou}
\affiliation{TRIUMF, 4004 Wesbrook Mall, Vancouver, British Columbia,
Canada, V6T 2A3} 

\author{R.~Poutissou} \affiliation{TRIUMF, 4004
Wesbrook Mall, Vancouver, British Columbia, Canada, V6T 2A3}

\author{G.~Redlinger} \altaffiliation{Present address: Brookhaven National Laboratory.}
\affiliation{TRIUMF, 4004 Wesbrook Mall, Vancouver, British Columbia,
Canada, V6T 2A3} 

\author{T.~Sasaki}
\affiliation{Research Center for Nuclear Physics, Osaka University,
10-1 Mihogaoka, Ibaraki, Osaka 567-0047, Japan} 

\author{T.~Sato} \affiliation{High Energy Accelerator
Research Organization (KEK), Oho, Tsukuba, Ibaraki 305-0801, Japan}

\author{T.~Shinkawa} \altaffiliation{Present address: Department of Applied
Physics, National Defense Academy, Yokosuka, Kanagawa 239-8686, Japan}
\affiliation{High Energy Accelerator Research Organization (KEK), Oho,
Tsukuba, Ibaraki 305-0801, Japan}

\author{F.C.~Shoemaker}
\affiliation{Joseph Henry Laboratories, Princeton University,
Princeton, New Jersey 08544} 

\author{A.J.S.~Smith}
\affiliation{Joseph Henry Laboratories, Princeton University,
Princeton, New Jersey 08544} 

\author{R. Soluk} \affiliation{Centre for
Subatomic Research, University of Alberta, Edmonton, Canada, T6G 2N5}

\author{J.R.~Stone} \affiliation{Joseph
Henry Laboratories, Princeton University, Princeton, New Jersey 08544}

\author{R.C.~Strand} \affiliation{Brookhaven National Laboratory,
Upton, New York 11973}

\author{S.~Sugimoto} \affiliation{High Energy
Accelerator Research Organization (KEK), Oho, Tsukuba, Ibaraki
305-0801, Japan} 

\author{C.~Witzig} \affiliation{Brookhaven National
Laboratory, Upton, New York 11973}

\author{Y.~Yoshimura}
\affiliation{High Energy Accelerator Research Organization (KEK), Oho,
Tsukuba, Ibaraki 305-0801, Japan}

\collaboration{E787 Collaboration}

\date{\today}

\begin{abstract}
We report the results of a search for the decay $K^+ \to \pi^+ \nu \bar \nu$ 
in the kinematic region with $\pi^+$ momentum $140 <  P < 195$  MeV/c using the data 
collected by the E787 experiment at BNL. No events were observed.
When combined with our previous search in this region, one candidate event with an expected background of
$1.22 \pm 0.24$ events results in a 90\% C.L. upper limit of $2.2 \times 10^{-9}$ on the branching
ratio of $K^+ \to \pi^+ \nu \bar \nu$. We also report improved limits on the rates of $K^+ \to \pi^+ X^0$ and
$K^+ \to \pi^+ X^1 X^2$ where $X^0, X^1, X^2$ are hypothetical, massless, long-lived neutral particles.
\end{abstract}

\maketitle

The decay $K^+ \to \pi^+ \nu \bar \nu$ is a flavor changing neutral
current  process which is highly suppressed at tree level by the Glashow-Iliopoulos-Maiani 
mechanism~\cite{GIM}; however, the breaking of  flavor symmetry, which results in  differences in the 
quark masses, allows this decay  to proceed at a very small rate at the loop level.
The large top quark mass results in dominance of the top quark contribution, 
making this decay very  sensitive  to the coupling of the top quark to the down 
quark, $V_{td}$, in the Cabibbo-Kobayashi-Maskawa  mixing matrix.
The branching ratio  for this rare decay has been measured to be
$(1.47^{+1.30}_{-0.89}) \times 10^{-10}$ based on the observation of three
events in the $\pi^+$ momentum phase space region 
$P > 211 {\rm ~MeV/c}$ (Region 1)~\cite{pnn1_e949}, 
using the data collected by the E787~\cite{pnn1,nim1,nim,ccd,ec,utc,td,pnn195} and E949~\cite{pnn1_e949,e949}
 experiments at the Alternating
Gradient Synchrotron (AGS) of Brookhaven National Laboratory. 
The standard model (SM) predicts the branching ratio for this decay mode to be
$(0.78 \pm 0.12) \times 10^{-10}$~\cite{smp1}.
Physics beyond the SM involving new heavy particles 
can interfere with the SM diagrams and alter the decay rate and the 
kinematic  spectrum~\cite{smp1,bsmp,bs}. It is therefore important 
to obtain  higher statistics for this decay and to extend the measurement 
to other regions of phase space. In an earlier paper~\cite{pnn2_96}, we reported 
the results from the kinematic search region below the $K^+ \to \pi^+ \pi^0 (K_{\pi2})$ 
peak with $\pi^+$ momentum, $140 <  P < 195$ (\rm MeV/c) (Region 2), 
using the data collected by the E787 experiment during the
1996 run. In this letter, we present  results for the same kinematic region 
from additional data collected during the 1997 run.

\begin{figure}
\begin{center}
\hspace{-1.2cm}
\epsfig{file=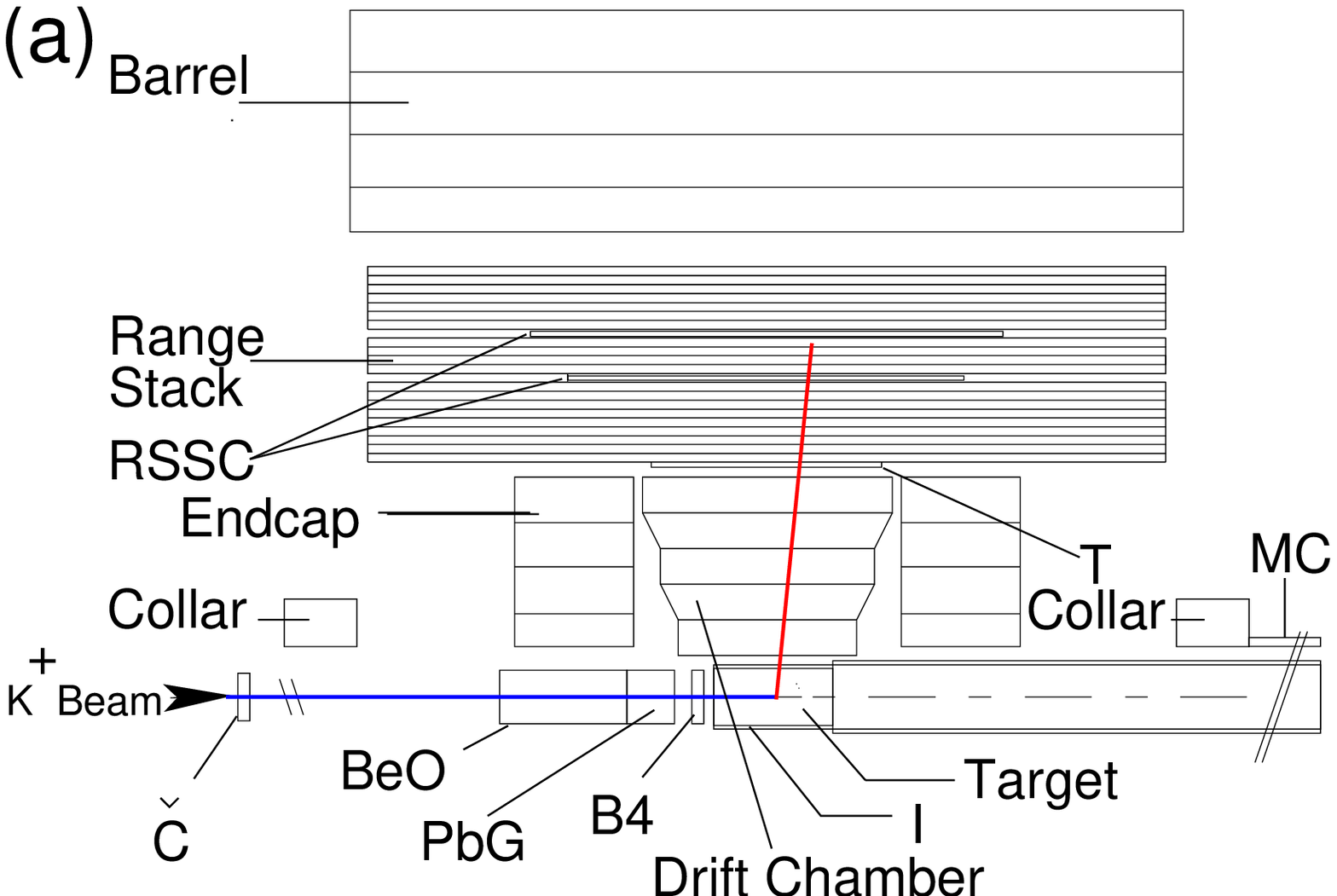,width=2.8in}
\end{center}
\end{figure}

\begin{figure}
\begin{center}
\vspace{-0.5cm}
\epsfig{file=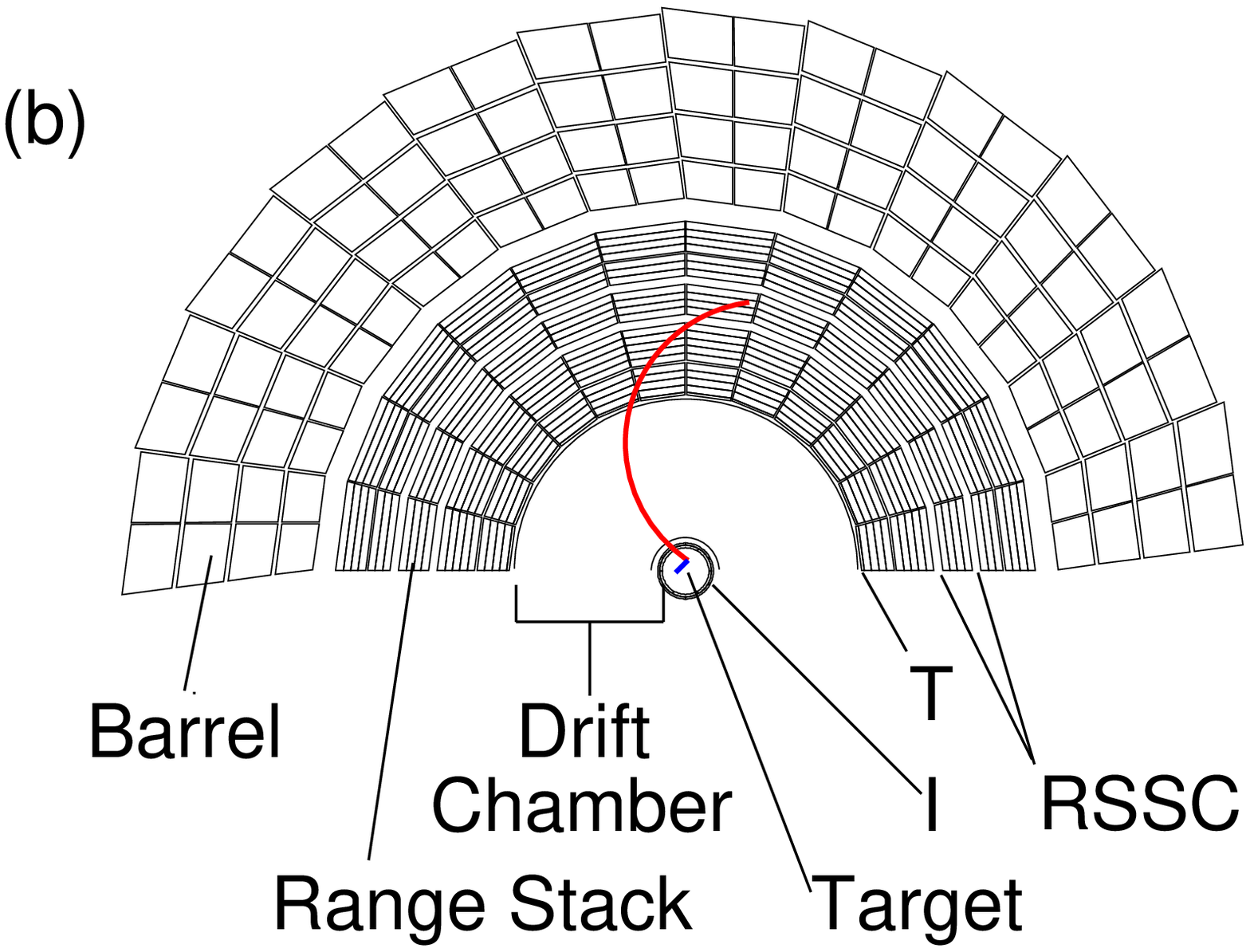,width=2.5in}
\end{center}
\caption{Schematic side (a) and end (b) views of the upper half of the E787 detector.
  \v{C} is the \v{C}erenkov counter, BeO is the passive degrader, PbG is the lead-glass detector, 
B4 is an energy loss counter, MC is the microcollar detector, I and T are the inner and
outer scintillation counters used in the trigger, RSSC are the Range Stack tracking chambers.}
\label{detector}
\end{figure}

\vspace{-0.05cm}
Event recognition for $K^+ \to \pi^+ \nu \bar \nu$  must rely on the detection of the incoming kaon
and the outgoing pion only,  since the two neutrinos are undetectable. 
Kaons were produced by  24 GeV protons from the 
AGS impinging on a 6~cm long platinum target.
The beam was  then transported, purified, and  momentum selected using 
electrostatic separators, two dipole magnets, several focussing 
magnets and collimators~\cite{nim}.
The resulting beam contained kaons and pions in the  ratio of 4:1.
The incoming beam particles were identified by a \v{C}erenkov detector and wire chambers
and were slowed down  by a passive BeO degrader and a lead-glass detector. The cylindrical lead-glass 
detector, which operated as a \v{C}erenkov light radiator 
(11.2 cm diameter and 10 cm long or 3.5 radiation lengths) was designed to detect 
incoming pions as well as electromagnetic
showers from  kaon decays, while being insensitive to the incident kaons.
\v{C}erenkov light from the lead-glass was collected from the sides of the cylinder
by a 1 cm thick Lucite sleeve coupled with silicone gel. The Lucite sleeve was glued to 16
azimuthally segmented trapezoidal lead-glass pieces. Each of these pieces was instrumented with a
fine-mesh photo-multiplier tube (PMT) which was  operated in the 1 T magnetic field.
The PMTs were located in a ring surrounding the BeO degrader and were read out by TDCs.  
Kaons were stopped in the target  consisting of  
413 $5 \rm mm \times 5 \rm mm \times 3.1 \rm m$ long plastic scintillating 
fibers, packed axially to form a 12~cm  diameter cylinder. 
Each  fiber was read out by a PMT. 
Gaps in the outer edges of the target were filled with 3.5 mm, 2 mm, and 1 mm fibers which were
connected to PMTs in groups. The PMTs were read out by ADCs, TDCs, and 500 MHz transient 
digitizers based on GaAs charge-coupled devices (CCDs)\cite{ccd}. The location, trajectory and  
momentum of the outgoing charged particles were measured by a drift chamber~\cite{utc}.
The range ($R$) and kinetic energy ($E$) of the charged particles were measured  
in a 21 layer Range Stack of plastic scintillator and the target.
The signals from the PMTs mounted on the Range Stack were recorded by ADCs and by
500 MHz transient digitizers (TDs)~\cite{td}. The TDs had the
ability to  record pulses for 6.4 microseconds after the event trigger, which enabled the observation
of the decay sequence, $\pi^+ \to \mu^+ \to e^+$ in the Range Stack. 
An electromagnetic calorimeter consisting of a 14-radiation length 
lead/scintillator barrel detector and 13.5-radiation length 
endcaps of undoped CsI crystals~\cite{ec} was used to veto photons.
Also, two collar counters composed of lead/scintillator with a total thickness of 4.6 radiation lengths  
and a microcollar detector composed of plastic scintillator fibers and lead foil stacked 
around the beamline with a radial thickness of 0.7 radiation length were used to detect photons which traveled at 
small angles relative to the beamline and thereby missed the barrel detector and the endcap detector.
Fig.~\ref{detector} shows  schematic side and end views of the E787 detector.

The data were obtained with a flux of $4.2 \times 10^6$ kaons per 1.6 s long spill (period
of 3.6 s) and with incident  kaon momenta of 710 MeV/c or 670 MeV/c~\cite{mom}. A multilevel 
$\pi^+ \nu \bar \nu$ trigger
ensured that a  kaon entered the target and decayed at rest, with an outgoing particle
identified as a pion by the $\pi^+ \to \mu^+$ decay sequence observed in the 
TD read-out of the Range Stack,  and with no other accompanying particles such as photons. 
In addition to the  $\pi^+ \nu \bar \nu$ trigger, several monitor triggers with variable pre-scale 
factors were  used to accumulate events from $K^+ \to \mu^+ \nu_{\mu}$ ($K_{\mu2}$) and $K_{\pi2}$ decays. A total of  
about 100 events per spill  were written to data storage devices.

To eliminate possible selection bias, a ``blind'' analysis technique was used, which
required the background sources to be identified {\it a priori} and the signal region to be
pre-defined. The events in the signal region were not counted or examined until the cuts and the background
estimates were final. At least two uncorrelated cuts 
with high rejection were designed for most of the backgrounds.
Each of these  cuts 
was independently reversed to create high statistics background samples to study the rejection power 
of the other cut. 
To minimize bias in the background measurements, the data were divided into one-third and two-third samples 
selected uniformly from the entire data set. Cuts were designed and optimized on the one-third sample and
the background was then measured on the two-third sample.

As discussed in~\cite{pnn2_96}, the background in Region 2 was found to be dominated by
$K_{\pi 2}$ events in which the charged pion underwent a nuclear interaction near 
the kaon decay vertex, most probably 
on a carbon nucleus in the target plastic scintillator. The scattering of the pion reduced its
kinetic energy into the signal region and removed the  directional correlation 
with the photons from the  $\pi^0$ decay, most of which would otherwise be detected by the 
barrel detector. The limited efficiency of the photon detectors along the  beam axis 
enhanced this background. During the 1997 run the lead-glass detector efficiency was 
improved in order to further suppress this background; however, most of the electromagnetic showers
detected in the lead-glass detector were also observed in the endcap detector so that no additional background
rejection was achieved. The improved efficiency of the lead-glass detector did, however, enable
an increase in the photon veto efficiency by about 10\% with respect to the 1996 data analysis.

The two sets of  analysis tools for suppressing the scattered  $K_{\pi2}$ background were the detection of photons from the 
$\pi^0$ decay and the identification of the $\pi^+$ scattering in the target. The signatures for scattering 
included kinks in the pattern of  target fibers attributed to the outgoing pion, tracks that did not point 
back to  the fiber in which the 
kaon decayed, energy deposits inconsistent with the ionization energy loss for a pion of the 
measured momentum,  extra deposited energy at the time of the outgoing pion in one of the target fibers 
traversed by the kaon.
The extra energy left by  pions in fibers traversed by the kaon was identified by examining the pulse
shapes recorded in the CCDs for each kaon fiber using a $\chi^2$ fit. 
Events were rejected in which an overlapping second pulse, coincident in time with the pion, was found to 
have energy larger than 1.5 MeV~\cite{cut_96}. 
Improved calibration of the target CCDs as well as this  higher energy threshold for 
the second pulse energy resulted in about 7\% more acceptance with the 
same level of background reduction in the  1997 data set as compared to the 1996 data set~\cite{bipul}.

\begin{table} 
\begin{center}
\begin{tabular}{|l|l|}
\hline
Process & Background \\
\hline 
$K^+ \to \pi^+ \pi^0$   &  $0.40 \pm 0.15$   \\
$K^+ \to \pi^+ \pi^0 \gamma$  & $0.006 \pm 0.002$  \\
$K_{\mu2\gamma} + K_{\mu3}$ & $0.009 \pm 0.009$ \\ 
Beam   & $0.033 \pm 0.030$  \\
$K^+ \to \pi^+ \pi^- e^+ \nu_e$ & $0.026 \pm 0.026$       \\
CEX & $0.013 \pm 0.013$   \\
\hline 
Total  & $0.49 \pm 0.16$  \\
\hline 
\end{tabular}
\end{center}
\caption{Expected number of background events  
in the search for $K^+\to \pi^+ \nu \bar\nu$ in Region 2 for the 1997 data 
set. The total background in the analysis of the 1996 data set was measured to be
$0.73 \pm 0.18$ in the same search region~\cite{pnn2_96}. 
The errors include both statistical and systematic uncertainties.}
\label{tab1}
\end{table}

\begin{table}
\begin{center}
\begin{tabular}{|l|r|}
\hline
Acceptance factors  &    \\
\hline 
$K^+$ stopping efficiency   & 0.708 \\
Time consistency cuts  & 0.672 \\ 
$K^+ \to \pi^+ \nu \bar\nu$ phase space  & 0.345   \\ 
Geometry   & 0.311 \\   
$\pi^+$ nucl. int. and decay in flight & 0.708   \\ 
Reconstruction efficiency & 0.943 \\    
Kinematic cuts & 0.697 \\     
$\pi - \mu - e$ decay chain & 0.490 \\ 
Beam and Target analysis  & 0.494  \\    
CCD acceptance  & 0.431 \\
Photon veto efficiency   & 0.390 \\ 
\hline 
Total acceptance & $9.7 \times 10^{-4}$   \\
\hline 
Total number of stopped $K^+$  & $0.61 \times 10^{12}$ \\
\hline 
Single Event Sensitivity & $1.69 \times 10^{-9}$ \\
\hline
\end{tabular}
\end{center}
\caption{Acceptance factors used in the measurement of 
$K^+\to \pi^+ \nu \bar\nu$  in Region 2. ``Time consistency cut'' requires
consistent times of the charged particles measured in different detector elements.
``Geometry'' includes the trigger efficiency and the efficiency of the first two layers of the Range Stack.
The uncertainty in the total acceptance is about 10\% which includes both statistical and systematic uncertainties.
Single event sensitivity is defined as the inverse of the product of the acceptance
and the total number of stopped kaons.}
\label{accept}
\end{table}

\begin{figure}
\begin{center}
\vspace{-2.0cm}
\epsfig{figure=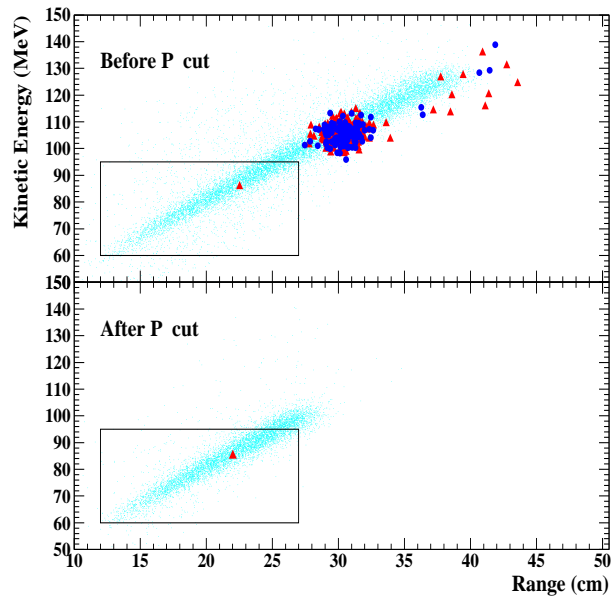,height=4.0in, width=3.5in}
\vspace{-1.3cm}
\end{center}
\hspace{-0.35cm}
\caption{ Kinetic energy (in MeV) versus range (in cm of plastic scintillator) distribution of  events 
that remained after all cuts except that on momentum (top), and after the momentum cut (bottom). 
The triangles represent  data  from the 1996 run  and the circles represent  data from the  1997 run. 
Events from Monte Carlo simulation of $K^+ \to \pi^+ \nu \bar\nu$
are represented by the light  dots.  The group of events around 108 MeV was due to   
$K_{\pi2}$ decays and the  events at higher energy were due
to $K_{\mu2}$ decays. 
All events except for the one in the signal region (shown by the rectangular box) 
were  eliminated by the $140 < P <195 {\rm ~MeV/c}$ cut on momentum. 
\vspace{-0.5cm} 
}
\label{rpeplt}
\end{figure}

The other backgrounds in the search for $K^+ \to \pi^+ \nu \bar \nu$ in Region 2 were 
$K^+ \to \pi^+ \pi^0 \gamma ~(K_{\pi2\gamma})$,
$K^+ \to \mu^+ \nu \gamma ~(K_{\mu2\gamma})$, 
$K^+ \to \mu^+ \nu \pi^0 ~(K_{\mu3})$,   
$K^+ \to \pi^+ \pi^- e^+ \nu_e $ ($K_{e4}$), scattered beam pions, 
and $K^+$ charge exchange (CEX) reactions resulting in 
decays $K^0_L \to \pi^+ l^- \bar\nu_l$, where $l = e$ or $\mu$. 
The contribution from these processes 
to the total background was small compared to 
scattered $K_{\pi2}$ decays.
The techniques used to measure  these backgrounds 
were discussed in~\cite{pnn2_96}.
Table~\ref{tab1} summarizes the contribution from each
background process. 
The uncertainty in the background estimate due to scattered $K_{\pi2}$ events was dominated by the
systematic uncertainty of the photon veto efficiency. This systematic uncertainty was estimated by measuring the
photon veto efficiency on several sets of scattered $K_{\pi2}$ events; each set was tagged by a different signature
for the scattering in the target~\cite{bipul}. The background estimates from the other processes were limited by statistics
of the data samples. Table~\ref{accept} shows the  acceptance factors and single event sensitivity. 
We gained about 6\% acceptance in the $K^+$ stopping efficiency due to the lower beam momentum 
during the 1997 run; another 14\% gain in acceptance was achieved by better 
optimization of cuts on the relative times of signals from the beam and target detectors.
However, we lost about 11\% in acceptance in the detection of the 
$\pi^+ \to \mu^+$ decay sequence due to a fault in the $\pi^+ \nu \bar \nu$ trigger  during the 1997 run. 
The effective branching ratio of the background is measured to be 
$0.486 \times 1.69 \times 10^{-9} = 8.2 \times 10^{-10}$, which
remains unchanged from the analysis of the 1996 data set~\cite{pnn2_96}. 
The analysis and detector improvements described in this paper led to an 
acceptance increase of about  27\% from the previous analysis~\cite{pnn2_96}.

Once the background estimates were finalized, 
the signal region (Region 2) was examined. No events were observed in the 1997 data set. 
The observation of one candidate event in the combined 1996-97 data set
is consistent with the total background estimate of $1.22 \pm 0.24$.
Fig.~\ref{rpeplt} shows the kinematics of the remaining events  
before and after the cut on momentum ($140 <P< 195$ MeV/c) for the combined data set. 
The event, reported in~\cite{pnn2_96}, has   $P=180.7 {\rm ~MeV/c}$, $R=22.1 {\rm ~cm}$, 
and $E=86.3 {\rm ~MeV}$ with a kaon decay time of 17.7 ns.

\begin{figure}
\begin{center}
\vspace{-0.1cm}
\epsfig{figure=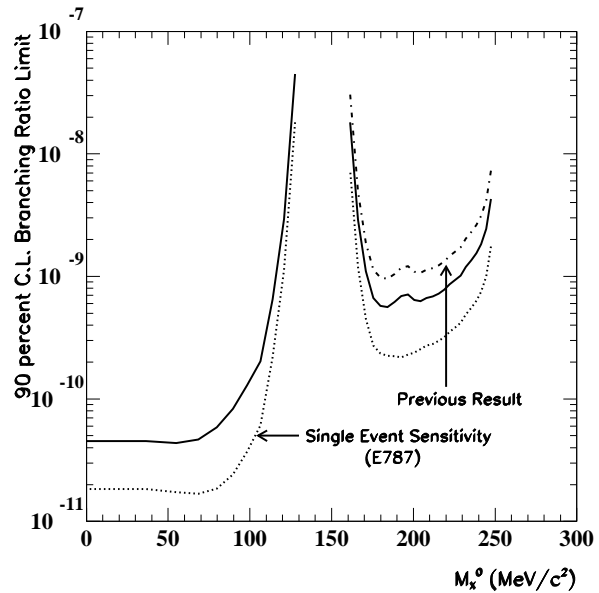, width=3.5in}
\vspace{-1.3cm}
\end{center}
\hspace{-0.35cm}
\caption{ The 90\% C.L. upper limit for $B(K^+ \to \pi^+ X^0$) as a 
function of $M_{X^0}$, where $M_{X^0}$ is  the mass of the recoiling system. 
The solid  line is from this analysis. ``Previous Result'' is from~\cite{pnn2_96}.  
The limit for $M_{X^0}< 140 {\rm ~MeV/c^2}$ is derived from the result for Region 1~\cite{pnn1}. 
The single event sensitivity as a function of  $M_{X^0}$ is shown by the dotted line.}
\label{xplot}
\end{figure}

The total numbers of $K^+$ stopped in the target were $1.12 \times 10^{12}$ and
$0.61 \times 10^{12}$ for the 1996 and 1997 data sets, respectively. 
Using the combined $K^+$ exposure,  the acceptances of $7.65 \times 10^{-4}$ from the 1996 data set~\cite{pnn2_96}
and that reported in Table~\ref{accept}, and  the observation of one 
event in  Region 2,  we calculated the
upper limit of $B(K^+\to \pi^+ \nu \bar\nu) < 2.2\times 10^{-9}$ (90\% C.L.)~\cite{feldman}. 
This result is a factor of two smaller than the result reported in~\cite{pnn2_96}.

For non-standard scalar and tensor  interactions in the process of $K^+ \to \pi^+ X^1 X^2$, where 
$X^1$ and $X^2$ are hypothetical, massless, long-lived neutral particles,
we set upper limits on their branching ratios at $2.7 \times 10^{-9}$ and $1.8 \times 10^{-9}$, 
respectively~(90\% C.L.)~\cite{feldman, scaler}.

This measurement is also sensitive to $K^+ \to \pi^+ X^0$, where $X^0$ is a hypothetical 
long-lived weakly interacting particle, or system of particles. Fig.~\ref{xplot} shows the 90\% C.L. upper 
limits on $B(K^+ \to \pi^+ X^0)$ together with the previous limit from \cite{pnn2_96}.

The search for $K^+ \to \pi^+ \nu \bar \nu$ in Region 2 was background limited and
another factor of five in background reduction is needed to achieve a signal to noise ratio of one
with the measured branching ratio in~\cite{pnn1_e949}.
New data from experiment E949~\cite{e949}, an upgraded version of E787, will provide further information in both
$K^+ \to \pi^+ \nu \bar \nu$ phase space regions.

\vspace{0.5cm}
\hspace{-0.35cm}{Acknowledgement} \\
We gratefully acknowledge the dedicated efforts of the technical staff 
supporting this experiment and the Brookhaven AGS Department. 
This research was supported in part by the U.S. Department of Energy 
under Contracts No. DE-AC0298CH10886, and grant 
DE-FG02-91ER40671, by the Ministry of Education, 
Culture, Sports, Science and Technology
of Japan through the Japan-US Cooperative Research 
Program in High Energy Physics and under the Grant-in-Aids for Scientific 
Research, encouragement of Young Scientists and for JSPS Fellows, and by 
the Natural  Sciences and Engineering Research Council and the National 
Research Council of Canada.

\end{document}